\documentclass{aastex61}
\usepackage{geometry}                
\usepackage{amsmath}
\usepackage{amssymb}

\usepackage{epstopdf}
\usepackage{graphicx}

\newcommand{\elp}{\ell'}
\newcommand{\eff}{{\mathcal F}}
\newcommand{\target}{{\mathcal T}}
\newcommand{\leak}{L}
\newcommand{\emp}{m'}
\newcommand{\eldp}{\ell{''}}
\newcommand{\emdp}{m{''}}
\newcommand{\emtp}{m{'''}}
\newcommand{\omdp}{\omega{''}}
\newcommand{\omtp}{\omega{'''}}

\newcommand{\kerne}{{{\mathcal{K}}}}

\begin{document}
\title{Measurement process and inversions using helioseismic normal-mode coupling}
\author{Shravan Hanasoge}
\altaffiliation{Department of Astronomy \& Astrophysics, Tata Institute of Fundamental Research, Mumbai 400005, India}
\altaffiliation{Center for Space Science, New York University Abu Dhabi, UAE}
\email{hanasoge@tifr.res.in}

\begin{abstract} 
Normal modes are coupled by the presence of perturbations in the Sun, providing a novel and under-appreciated helioseismic technique with which to image the solar interior. The process of measuring coupling between normal modes is straightforward, much more so when compared with other prevalent helioseismic techniques. The theoretical framework to interpret these measurements is well developed with the caveat that it applies only in the case where the entire surface of the Sun is observed. In practice however, the limited visibility of the Sun and line-of-sight related effects diminish the resolution of the technique. Here, we compute realistic sensitivities of normal-mode coupling measurements to flows in the solar interior and describe how to mitigate the sometimes-overwhelming effect of leakage. The importance of being able to isolate individual spherical harmonics and observe the full Sun, to which future solar observatories may aspire, is thus highlighted in our results. In the latter part of the article, we describe the noise model for the variance of coupling coefficients, a critical component to the process of inference.
\end{abstract}


\section{Introduction}\label{intro}
Global-mode helioseismology has led to robust inferences of axisymmetric properties of the Sun such as its structure (as a function of radius) and rotation \citep[with radius and latitude; e.g.][]{jcd02}. However global-mode frequencies possess very little sensitivity to meridional circulation and non-axisymmetric features such as convection. Nevertheless, frequency series' of mode harmonics, i.e. $\phi_{\ell m}(\omega)$, where $\omega$ is temporal frequency, $\ell$ is spherical harmonic degree, $m$ the azimuthal order and $\phi$ the observed line-of-sight projected surface velocity, contain significant information that has not been fully exploited. In global-mode analyses, the auto-correlation quantity $\phi_{\ell m}^*\,\phi_{\ell m}$ is the primary measurement. Whereas, the more general quantity $\phi_{\ell' m'}(\omega')\,\phi^*_{\ell m}(\omega)$, which is the wavefield correlation in Fourier domain, contains phase information (unlike global-mode frequencies), providing insight into non-axisymmetric and time-varying features in the solar interior. There have been attempts to infer meridional circulation and convection using these measurements \citep[e.g.][]{schad, woodard13, woodard14,woodard16} but the technique still needs to be refined and understood better in order to place faith in subsequent inferences. Based on the original work by \citet{lavely92} and techniques of terrestrial seismology \citep{DT98}, \citet{hanasoge17_etal} and \citet{hanasoge17} computed kernels for flows and magnetic fields respectively \citep[also see][]{cutler}. However, these kernels were predicated on being able to observe the solar surface in its entirety and thus do not capture systematical errors associated with realistic observations of the Sun.

The premise of normal-mode coupling is as follows. Consider the eigenfunctions of a reference model of the Sun, i.e. a structure model that is non-rotating, non-convecting, non-magnetic and radially stratified \citep[e.g. model S, ][]{jcd}. These eigenfunctions are orthogonal and form a complete basis \citep{jcd_notes}. Assuming the eigenfunctions of the real Sun to be slightly perturbed forms of the reference eigenfunctions, we may express the former as linear combinations of the latter. Thus the eigenfunction of a given mode in the real Sun is actually a mixture of eigenfunctions of the reference Sun, and the modes of the real Sun are said to be `coupled' with respect to the reference. However since we only observe the surface, it is not possible to measure the eigenfunction, a 3-D quantity, in the Sun. We thus use the measurement $\phi_{\ell' m'}(\omega')\,\phi^*_{\ell m}(\omega)$ as a proxy for the extent of coupling.
Because we only observe half the Sun, and in practice a third, owing to limb-darkening effects, it is not possible to isolate modes in terms of spherical harmonics. 
Thus, even in the reference model, the quantity $\phi_{\ell' m'}(\omega')\,\phi^*_{\ell m}(\omega)$ is non-zero owing to effects such as leakage, realization noise (waves in the Sun are stochastically excited), and other systematic effects. Mode coupling induced by systematical effects must therefore be computed with great care since they contribute a dominant fraction of measured mode correlations \citep[for an equivalent approach in time-distance helioseismology, see e.g.][]{gizon_04}. 

The uncertainty principle indicates that windowing in the spatial domain results in a broadening in the spherical-harmonic domain, causing mode signal to ``leak'' from one spherical harmonic to neighbouring channels. To characterize this effect, \citet{schou94} introduced the so-called leakage matrix $L^{\ell m}_{\ell' m'}$, where $(\ell,m)$ and $(\ell',m')$ are two modes. Leakage between modes diminishes rapidly with increasing $|\ell-\ell'|$ and $|m-m'|$. One consequence of leakage is that the wavefield correlation $\phi_{\ell' m'}^*(\omega')\,\phi_{\ell m}(\omega)$ actually measures the coupling between a number of modes (and not just between $\ell,m$ and $\ell',m'$), the details of which depend in complicated ways on the resonant frequencies of the modes and specific wavenumbers in question. Mode leakage in turn has the effect of garbling the sensitivity of the measurement to underlying perturbations. In other words, the inability to isolate a specific modal harmonic in the observations translates to  inferences of the perturbation, i.e. it is impossible to retrieve individual wavenumbers. The important issues are therefore of quantifying, accounting for and mitigating this effect.

{The present analysis relies on Doppler-velocity measurements of the surface - thus the observable is the line-of-sight projected velocity field.} The measured signal at any given point in the frequency-wavenumber spectrum comprises contributions from modes at all temporal frequencies (i.e. all radial orders) at that specific spatial frequency. This is due to the fact that we cannot observe the internal layers of the Sun and isolate specific radial orders. However, what makes it feasible to identify (approximately, at any rate) individual radial orders is that mode power, which is well approximated as a Lorentzian, falls rapidly away from resonance. Thus close to resonance (within a couple of mode linewidths), the observed signal is dominated by the mode at that radial order, and we therefore associate signal within that region of the spectrum with that radial order. Because space-based instruments contain essentially no temporal gaps, observations may be well isolated in temporal frequency. Using this reasoning, we use temporal frequency as a means to discern radial order, and stay within a few linewidths of resonance.

The helioseismic wave equation \citep[e.g.][]{jcd02,gizon2010} is linear and as a consequence, the wavefield at different temporal frequency channels may be treated independently. However, convection comprises non-axisymmetric time-varying features, introducing a convolution in the frequency domain and thereby mixing frequency channels. The analysis discussed in e.g. \citet{woodard16} and \citet{hanasoge17_etal} address this issue by considering mode-correlation measurements at different frequencies across a mode. As discussed earlier though, these measurements need to made very close to resonance. One consequence therefore of limiting the frequency interval of study (to avoid straying too far from resonance) is that it is not possible to study convective features at short lifetimes (the shorter the lifetime, the larger the associated frequency).

In this article, we incorporate the systematical issues associated with leakage and derive corresponding sensitivity functions for flows. We also construct a theoretical model for noise levels that are expected in these measurements.


\section{The fully visible Sun}
The theory describing the connection between wavefield correlations and perturbations is detailed in \citet{woodard16} and \citet{hanasoge17_etal}. To summarize the development, the sensitivity of coupled modes $(\ell, m, n)$ and $(\ell', m', n)$ due to flows in the interior is given by
\begin{equation}
\phi^{\omega+\sigma}_{\ell' m'}\,\phi^{\omega*}_{\ell m} = -2\omega(N_{\ell'}\,R^{\omega*}_{\ell m}\,|R^{\omega+\sigma}_{\ell' m'}|^2 + N_{\ell}\,R^{\omega+\sigma}_{\ell' m'}\,|R^{\omega}_{\ell m}|^2) \sum_{s,t} \int_\odot dr\,[iu_{st}(r, \sigma)\,K_{st}^u(r) +
w_{st}(r, \sigma)\,K_{st}^w(r)],\label{phiphi}
\end{equation}
where the dependence on temporal frequency $\omega$ is written now as a superscript, i.e. $\phi^\omega_{\ell m} = \phi_{\ell m}(\omega)$, $\sigma$ is the inverse timescale associated with the flow, $r$ is radius, $u_{st}(r,\sigma)$ denotes mass-conserving poloidal flow \citep[see Eqs.~61 and~62 of][]{hanasoge17_etal} at harmonic degree $s$ and azimuthal order $t$ and $w_{st}(r, \sigma)$ the toroidal flow component (mass conserving by construction). The mode frequency and amplitude details are captured by terms $N_{\ell}$ and $R^\omega_{\ell m}$, the mode-normalization constant and Lorentzian respectively, where
\begin{equation}
R_{\ell m}^{\omega}=\frac{1}{(\omega_{n\ell m} - i\Gamma_{n\ell}/2)^2 - \omega^2} \approx \frac{1}{2\omega_{n\ell m}}\frac{1}{\omega_{n\ell m} - i\Gamma_{n\ell}/2 - \omega},\label{Rdef}
\end{equation}
and $\Gamma_{n\ell}$ is the damping rate. The sensitivity kernels $K_{st}^u$ and $K_{st}^w$ are actually dependent on $\ell, m$ and $\ell',m'$ but to reduce notational burden, we do not explicitly state them. These kernels may be simplified significantly using asymptotic expressions derived in \citet{woodard14} and Appendix~\ref{asympsec}
\begin{eqnarray}
K_{st}^u \approx \gamma^{\ell' s \ell}_{t m}\,g_{\ell' - \ell, s}\, \kerne_{n\ell}(r),\\
K_{st}^w \approx \gamma^{\ell' s \ell}_{t m}\,f_{\ell' - \ell, s}\, \kerne_{n\ell}(r),
\end{eqnarray}
where we slightly tweak the notation introduced by \citet{woodard14} for the symbol $\gamma$, which is defined thus
\begin{equation}
\gamma^{\ell' s \ell}_{t m} = (-1)^{m+t}\,\sqrt{2s+1}{ 
\begin{pmatrix} 
{\ell'} & s & \ell \cr
{-(m+t)} & t & m 
\end{pmatrix}}.\label{C-B}
\end{equation}
The term in the parenthesis of equation~(\ref{C-B}), the Clebsch-Gordan coefficient, is non-zero only if the second row sums to zero (and hence there are only two independent parameters in the second row) and $|m+t| \le \ell', |t| \le s, |m| \le \ell, |\ell'-\ell | \le s, |\ell - s| \le \ell'$ and $|\ell'-s| \le \ell$. The symbols $f$ and $g$ are given by
\begin{eqnarray}
f_{\ell'-\ell,s} = (-1)^{(s + \ell'-\ell-1)/2}\frac{(s-\ell'+\ell)!!\,(s+\ell'-\ell)!!}{\sqrt{(s-\ell'+\ell)!\,(s+\ell'-\ell)!}}\,\,\,\,\,\,\,\, ({\rm for\,odd}\, s+\ell'-\ell),\label{feq}\\
g_{\ell'-\ell,s} = i\, (-1)^{(s + \ell'-\ell)/2}\,(\ell' - \ell)\,\frac{(s-\ell'+\ell-1)!!\,(s+\ell'-\ell-1)!!}{\sqrt{(s-\ell'+\ell)!\,(s+\ell'-\ell)!}}\,\,\,\,\,\,\,\, ({\rm for\,even}\, s+\ell'-\ell),\label{geq}
\end{eqnarray}
where we emphasise that $g$ absorbs the factor of $i$ from equation~(\ref{phiphi}) and is therefore complex.
The term $f_{\ell'-\ell,s}$ in equation~(\ref{feq}) is defined to be non-zero only for odd $s + \ell' - \ell$ whereas $g_{\ell'-\ell,s}$ (Eq.~[\ref{geq}]) is defined to be non-zero only for even $s + \ell' - \ell$ and $\ell'\neq\ell$. Finally the term $\kerne_{n\ell}(r)$ is given by
\begin{equation}
\kerne_{n\ell}(r) = \frac{(-1)^\ell}{\sqrt{2\pi}} \ell^\frac{3}{2} \rho r [ U_{n\ell}^2 + \ell(\ell+1) V_{n\ell}^2 ],
\end{equation}
where $U_{n\ell}(r)$ and $V_{n\ell}(r)$ are the radial and horizontal eigenfunctions for the $(n,\ell)$ mode \citep[e.g., see][]{jcd_notes} and $\rho$ is density.

We now consider measurements of mode coupling where $\ell = \ell'$ and $s$ is odd \citep[as studied, e.g. by][]{woodard16}. The relationship between measurement and flow is given by
\begin{equation}
\phi^{\omega+\sigma}_{\ell m+t}\,\phi^{\omega*}_{\ell m} = H^\sigma_{\ell \ell m t}(\omega) \sum_{s} \gamma^{\ell s \ell}_{t m}\,\int_\odot dr\,[g_{0, s} \,u_{st}(r, \sigma)\,\kerne_{n\ell}(r) +
f_{0, s} \,w_{st}(r, \sigma)\,\kerne_{n\ell}(r)].
\end{equation}
From equation~(\ref{phiphi}), we denote the term that captures the mode power as a function of frequency through another symbol, $H$,
\begin{equation}
H^{\sigma}_{\ell' \ell mt}(\omega) = -2\omega(N_{\ell'}\,R^{\omega*}_{\ell m}\,|R^{\omega+\sigma}_{\ell' m+t}|^2 + N_{\ell}\,R^{\omega+\sigma}_{\ell' m+t}\,|R^{\omega}_{\ell m}|^2),
\end{equation}
facilitating future analyses.
From equation~(\ref{geq}), we see that $g_{0,s} = 0$ and therefore the sensitivity of the measurement is given by
\begin{equation}
\phi^{\omega+\sigma}_{\ell m+t}\,\phi^{\omega*}_{\ell m} = H^\sigma_{\ell \ell mt}(\omega) \sum_{s} f_{0, s} \,\gamma^{\ell s \ell}_{t m}\,\int_\odot dr\,
w_{st}(r, \sigma)\,\kerne_{n\ell}(r),
\end{equation}
which is an elegant and simple expression. To aid analysis of observed data, \citet{woodard16} introduce $b$-coefficients that are defined through the following reciprocal relationships,
\begin{equation}
\phi^{\omega+\sigma}_{\ell m+t}\,\phi^{\omega*}_{\ell m} = \sum_{s} \gamma^{\ell s \ell}_{t m}\,H^\sigma_{\ell \ell mt}(\omega)\, b^\sigma_{st}(n,\ell),\label{b1def}
\end{equation}
and the linear-least-squares approximation to the $b$ coefficient is given by \citep[adopted, for instance, by][]{woodard16},
\begin{equation}
b^\sigma_{st}(n,\ell) = \frac{\sum_{m,\omega} H^{\sigma*}_{\ell \ell mt}\,\gamma^{\ell s \ell}_{t m}\, \phi^{\omega+\sigma}_{\ell m+t}\,\phi^{\omega*}_{\ell m}}{\sum_{m,\omega} |H^{\sigma}_{\ell \ell mt}\,\gamma^{\ell s \ell}_{t m}|^2},\label{bdef}
\end{equation}
where, in order to ensure that the wavefield is sampled closed to resonance, the frequency interval over which the sum is carried out must satisfy the criterion 
\begin{equation}\omega \in (\omega_{n\ell m} - \Gamma_{nl}, \omega_{n\ell m} + \Gamma_{nl})\,\,\,\, or\,\,\, \omega \in (\omega_{n\ell m+t} - \Gamma_{nl}  - \sigma, \omega_{n\ell m+t} + \Gamma_{nl} - \sigma).\label{freqintervals} \end{equation} 
Note that $\omega > 0$ in both these intervals. 
With this definition of the $b$ coefficient, we have
\begin{eqnarray}
g_{0,s}\int_\odot dr\,u_{st}(r, \sigma)\,\kerne_{\ell n}(r) + f_{0,s}\int_\odot dr\,w_{st}(r, \sigma)\,\kerne_{\ell n}(r)\, = b^\sigma_{st}(\ell, n),\label{btoflow}
\end{eqnarray}
with the poloidal term $u_{st}(r,\sigma)$ written only for the sake of completeness (Eq.~[\ref{geq}] gives us $g_{0,s}=0$).
The flow system $w_{st}$ is non-axisymmetric, representing a temporally and  spatially fluctuating quantity. In order to stabilise the process of inference, we consider the power spectrum of $w$, i.e. the structure function of turbulence, which is axisymmetric and therefore a more stable quantity. As an aside, we note that because the flow field is real in the spatio-temporal domain, $b$ coefficients obey the parity relationship $b^{\sigma*}_{st} = (-1)^t\,b^{-\sigma}_{s,-t}$.

\subsection{Interpretation using wavefield correlations in ideal case}\label{ideal.inverse}
We combine the measurements $b_{st}$ over various $\ell$ and $n$ so that the corresponding sum of kernels is focused around a localized region in radius. This gives
\begin{eqnarray}
\int_\odot dr\,w_{st}(r, \sigma)\,f_{0,s}\sum_{\ell,n} \alpha^{\ell n}\,\kerne_{\ell n}(r)\, = \sum_{\ell,n} \alpha^{\ell n}\, b^\sigma_{st}(\ell, n),\\
\int_\odot dr\,w_{st}(r, \sigma)\,\eff_{s}(r)\, = \beta^\sigma_{st},
\end{eqnarray}
where $\eff_s(r)$ is a radially focused function (such as a Gaussian centered around a given radius) and $\beta^\sigma_{st}$ is the combined sum over $b$ coefficients measured at various $(n,\ell)$, weighted by $\alpha$. Now we multiply both sides by the complex conjugate of $\beta$ and sum over $t$ to obtain
\begin{equation}
\int_\odot dr\,dr'\,\sum_t  w_{st}(r, \sigma)\,w^*_{st}(r', \sigma)\,\eff_{s}(r)\,\eff_{s}(r') = \sum_t |\beta^\sigma_{st}|^2.
\end{equation}
If $\eff_s(r)$ is generally well focused around some radius $r_0$, we may approximate $\eff_s(r)\,\eff_s(r')\approx \delta(r-r_0)\,\delta(r-r')$,
\begin{equation}
\int_\odot dr\,\sum_t  |w_{st}(r, \sigma)|^2\,\delta(r-r_0) = \sum_t |\beta^\sigma_{st}|^2.
\end{equation}
and we obtain
\begin{equation}
\sum_t  |w_{st}(r_0, \sigma)|^2 = {\mathcal P}_s(r_0) = \sum_t |\beta^\sigma_{st}|^2.
\end{equation}
Note that we have not stated considerations such as the tradeoff between sharpness of localisation in depth and the noise level \citep[e.g.][]{pijpers} but these need also to be taken into account when deriving the weighted sum $\beta^\sigma_{st}$. 
Thus in the scenario where we are able to isolate individual modes in the spherical harmonic domain, we can obtain accurate estimates of convective properties in the interior. In the next section, we consider the realistic situation, where leakage and other effects play an important role.

\section{The partially visible Sun}
As described in Section~\ref{intro}, we only observe a third of the Sun's surface, diminishing our ability to isolate individual modes. To take this into account, we use the leakage matrix $\leak^{\ell' m'}_{\ell m}$, whose action is described thus,
\begin{equation}
\phi_{\ell m}(\omega) = \sum_{\ell' m'} \leak^{\ell' m'}_{\ell m}\,a_{\ell' m'}(\omega),\label{leakrel}
\end{equation}
where $a_{\ell' m'}(\omega)$ denotes the `true' oscillations of mode $(\ell', m', n)$, i.e. for $\omega \approx \omega_{\ell' m' n}$ and
$\phi_{\ell m}(\omega)$ the observed oscillations. Note that the leakage matrix is a function of radial order but we do not explicitly state $n$ since we use frequency as a proxy for radial order and also $\omega$ always lies within a few linewidths of resonance (Eq.~[\ref{freqintervals}]). In the calculations that follow, we use a standard leakage matrix that takes into account line-of-sight projection, limb darkening and comprises both horizontal and radial terms. We note that because we must now distinguish between `true' and observed oscillations, we will need to correspondingly consider `true' and observed $b$ coefficients, denoted by $B^\sigma_{st}$ and $b^{\sigma}_{st}$ respectively.

Thus, given a true state of mode coupling, instrumental and other systematics corrupt our retrieved estimates of $b$ coefficients. There is no way to correct for this since the observations are incomplete. The best we can hope to accomplish is to model the systematic effects, mitigate them and appreciate the actual information content of our measurements. Applying the definition of the $b$ coefficient from equation~(\ref{bdef}) and keeping in mind that we are now dealing with observations, we obtain
%
%
%
\begin{eqnarray}
B^\sigma_{st}(n,\ell) = N^{\sigma}_{\ell st}\sum_{m,\omega}\, \gamma^{\ell s \ell}_{t m}\,H^{\sigma*}_{\ell\ell mt}(\omega)\, \phi^*_{\ell m}(\omega)\, \phi_{\ell, m+t}(\omega+\sigma),\label{Bcoef}\\
N^{\sigma}_{n\ell st} = \frac{1}{\sum_{m,\omega} |H^{\sigma}_{\ell \ell mt}\,\gamma^{\ell s \ell}_{t m}|^2},
\end{eqnarray}
where the summation over frequency follows the interval described in equation~(\ref{freqintervals}). 
Introducing the leakage relation~(\ref{leakrel}) into equation~(\ref{Bcoef}), we obtain a significantly bulkier expression,
\begin{equation}
B^\sigma_{st} = 
N^{\sigma}_{\ell st}\,\sum_{\elp,\eldp,m,\emp,t',\omega}{}\leak^{\elp\emp}_{\ell m}\,{}\leak^{\eldp\emp+t'}_{\ell m+t}\,\gamma^{\ell s \ell}_{t m}\, H^{\sigma*}_{\ell\ell mt}\,   a^{\omega*}_{\elp\emp}\,\, a^{\omega+\sigma}_{\eldp\emp + t'},\label{Bcloser}
\end{equation}
where $a$ denotes the true wavefield. Substituting equation~(\ref{b1def}) that relates the true wavefield correlation to the corresponding true $b$ coefficients,
\begin{equation}
a^{\omega*}_{\elp\emp}\,\, a^{\omega+\sigma}_{\eldp \emp+t'}= \sum_{s't'}\gamma^{\eldp s'\elp}_{t' m'}\, H^{\sigma}_{\elp\eldp m' t'}\, b^\sigma_{s't'}(\ell{'},\ell{''}).\label{acoef}
\end{equation}
Replacing the wavefield correlation on the right-hand side of equation~(\ref{Bcloser}) with~(\ref{acoef}), we obtain
\begin{equation}
B^\sigma_{st}= N^{\sigma}_{\ell st}\,\sum_{\elp,\eldp,m,\emp,s',t',\omega}{}\leak^{\elp\emp}_{\ell m}\,{}\leak^{\eldp\emp+t'}_{\ell m+t}\, \gamma^{\ell s\ell}_{t m}\, H^{\sigma*}_{\ell\ell mt}\,\gamma^{\eldp s'\elp}_{t' m'}\, H^{\sigma}_{\elp\eldp m' t'}\, b^\sigma_{s't'}(\ell{'},\ell{''}).\label{Bnext}
\end{equation}
Equation~(\ref{Bnext}) tells us that the observed $B$ coefficient is a mixture of $b$-coefficients at several spatial wavenumbers, and it may therefore not be possible to cleanly separate features at different wavenumbers. The degree of this mixing is however mitigated by the leakage terms and $H^\sigma_{\elp\eldp m' t}$, which represents Lorentzians centred around the mode frequencies $\omega_{n\ell m}$ and $\omega_{n\ell m+t}$. As distances $|\eldp - \ell|, |\elp - \ell|, |m'+t' -m|$ and $|m' -m|$ grow, the $H$ factors and leakage matrices fall ever more rapidly. It now bears remembering that the true $b^\sigma_{s't'}(\elp,\eldp)$ coefficient no longer contains contributions only from modes of spherical-harmonic degree $\ell$ - rather, it is from modes $\elp$ and $\eldp$ with the possibility that $\elp\neq\eldp$. This implies that the measured $B$ coefficient is sensitive to both toroidal {\it and} poloidal flows (see Eqs.~[\ref{feq}] and~[\ref{geq}]).
Substituting equation~(\ref{btoflow}) into~(\ref{Bnext}), we obtain
\begin{eqnarray}
B^\sigma_{st}  = N^{\sigma}_{\ell st}\,\sum_{\elp,\eldp,m,\emp,s',t',\omega}{}\leak^{\elp\emp}_{\ell m}\,{}\leak^{\eldp\emp+t'}_{\ell m+t}\,\gamma^{\ell s \ell}_{t m}\,  H^{\sigma*}_{\ell\ell mt}\,\gamma^{\eldp s'\elp}_{t' m'}\, H^{\sigma}_{\elp\eldp m' t'} \,\times \nonumber\\
\left[ g_{\ell{''}-\ell{'},s'}\,\int_\odot dr\,  u_{s't'}(r,\sigma)\, \kerne_{n\ell}(r) +  f_{\ell{''}-\ell{'},s'} \,\int_\odot dr\, w_{s't'}(r,\sigma)\, \kerne_{n\ell}(r)\right],
\end{eqnarray}
where we recall that $|\elp - \eldp| \ll \ell, s'$.
Combining mode-coupling measurements at different radial orders and harmonic degrees, i.e. multiplying both sides by $\alpha^{n\ell}$ as in Section~\ref{ideal.inverse}, and summing over $n$ and $\ell$, we obtain an elegant result
\begin{eqnarray}
\beta^\sigma_{st}(\sigma) = \sum_{n\ell} \alpha^{n\ell}\,B^\sigma_{st} =  \int_\odot dr\,\sum_{s',t'} P^{s't'}_{st}(r,\sigma)  u_{s't'}(r,\sigma) + T^{s't'}_{st}(r,\sigma)  w_{s't'}(r,\sigma),\label{betaPT}
\end{eqnarray}
where
\begin{eqnarray}
&&P^{s't'}_{st}(r,\sigma) = \sum_{\ell,n} \alpha^{n\ell}\, \Pi^{s't'}_{st}(n,\ell,\sigma),\label{p.eq}\\
&&\Pi^{s't'}_{st}(n,\ell,\sigma)=N^{\sigma}_{\ell st}\,\kerne_{n\ell}(r)\,\sum_{\elp,\eldp,m,\emp,\omega}g_{\eldp-\elp,s}\, \leak^{\elp\emp}_{\ell m}\,\leak^{\eldp\emp+t'}_{\ell m+t}\,\gamma^{\ell s \ell}_{t m}\,H^{\sigma*}_{\ell\ell mt}\,\gamma^{\eldp s'\elp}_{t' m'}\, H^{\sigma}_{\elp\eldp m' t'},
\end{eqnarray}

\begin{eqnarray}
&&T^{s't'}_{st}(r,\sigma) = \sum_{\ell,n} \alpha^{n\ell}\, \Theta^{s't'}_{st}(n,\ell,\sigma).\label{q.eq}\\
&&\Theta^{s't'}_{st}(n,\ell,\sigma)=N^{\sigma}_{\ell st}\,\kerne_{n\ell}(r)\,\sum_{\elp,\eldp,m,\emp,\omega} f_{\eldp-\elp,s}\, \leak^{\elp\emp}_{\ell m}\,\leak^{\eldp\emp+t'}_{\ell m+t}\,\gamma^{\ell s \ell}_{t m}\,H^{\sigma*}_{\ell\ell mt}\,\gamma^{\eldp s'\elp}_{t' m'}\, H^{\sigma}_{\elp\eldp m' t'}.
\end{eqnarray}
Note that, in principle, $P$ and $T$ are complex functions, mixing both amplitude and phases of the underlying turbulent velocities. In practice, the imaginary component is smaller by at least 3 orders in magnitude so we may treat $P$ and $T$ as real. This is primarily due to the fact that the leakage matrix is purely real, and may not be fully accounting for all systematical issues (see also Section~\ref{leaksec}, related to the leakage matrix). The terms within square brackets are highlighted since they denote
functions purely of $\ell,n,s,s',t,$ and $t'$. If the entire Sun were visible, the highlighted terms would be $\propto \delta_{s,s'}\,\delta_{t,t'}$.

\section{Subtractive Optimally Localized Averaging (SOLA)}\label{SOLA}
SOLA is a useful technique with which to incorporate a variety of constraints while limiting the noise level on the eventual inference \citep{pijpers}. The combination of measurements results in a corresponding combination of the associated kernels, termed here as the ``averaging kernel". In the present problem, we require the averaging kernel sensitive to a given flow wavenumber $(s,t)$ be localized around a given depth, that it possess no surface tail. Additionally, the problem of  leakage, described earlier, results in an inability to isolate a single $(s,t)$ mode. Consequently, we penalize kernels sensitive to flows at neighbouring wavenumbers $(s', t') \neq (s,t)$. These three constraints form the SOLA problem relevant to mode coupling. The translation of this into a mathematical form requires solving the following minimization problem
\begin{equation}
\chi_{st} = \frac{1}{2}\int_\odot dr \left[T^{st}_{st} - \target(r)\right]^2 + \frac{\lambda}{2}\int_\odot dr\,\sum_{s',t'}\left(T^{s't'}_{st}\right)^2  + \frac{\nu}{2}\int_\odot dr f(r)\,\sum_{s',t'}\left(T^{s't'}_{st}\right)^2,\label{penalties} 
\end{equation}
where $\target(r)$ is the target function, $T^{s't'}_{st}$ is defined in equation~(\ref{q.eq}) and depends on $\alpha^{n\ell}$, which are parameters that are used to combine the modes. The function $f(r)$ is applied to penalise surface tails of the the averaging kernel $T^{st}_{st}$ (see appendix~\ref{surfconstraint} for a description of function $f(r)$). With the noise model described in Section~\ref{noise.model}, it is also possible to add a regularization term to control the eventual noise in the inferences. For the moment, we do not include this term but a description of it may be found in e.g. \citet{pijpers}.

The sum over $(s',t') \neq (s,t)$ penalizes the sensitivity of the measurement combination to flows at neighbouring wavenumbers (undesired). The parameters $\lambda$ and $\nu$ are for regularization and must be determined through trial and error (or for instance, the optimum choice is commonly associated with the knee of penalty function, when plotted as a function of the noise regularization parameter). 
To obtain optimal coefficients $\{\alpha\}$, we differentiate $\chi_{st}$ with respect to a specific $\alpha^{n\ell}$ and set it to zero,
\begin{eqnarray}
&&\frac{\partial\chi_{st}}{\alpha^{n\ell}} = \sum_{n'\ell'}\left[\int_\odot dr \, \Theta^{st}_{st}(r;n,\ell,\sigma)\,\Theta^{st}_{st}(r;n',\ell',\sigma) \right.\nonumber\\ &&\left.+ [\lambda + \nu f(r)]\,\sum_{s',t'} \Theta^{s't'}_{st}(r;n,\ell,\sigma)\,\Theta^{s't'}_{st}(r;n',\ell',\sigma) \right]\alpha^{n'\ell'}
=\int_\odot dr\, \Theta^{st}_{st}(r;n,\ell,\sigma)\, \target(r),
\end{eqnarray}
where $(s',t')\neq(s,t)$. Defining matrix $A$ whose matrix elements are given by
\begin{equation}
A_{n\ell,n'\ell'} =  \int_\odot dr\, \Theta^{st}_{st}(r;n,\ell,\sigma)\,\Theta^{st}_{st}(r;n',\ell',\sigma) +[\lambda + \nu f(r)]\,\sum_{s',t'} \Theta^{s't'}_{st}(r;n,\ell,\sigma)\,\Theta^{s't'}_{st}(r;n',\ell',\sigma),\label{matA}
\end{equation} 
and the column vector $\{R\}$, 
\begin{equation}
R_{n\ell}=\int_\odot dr\, \Theta^{st}_{st}(r;n,\ell,\sigma)\, \target(r),\label{invpr}
\end{equation}
we arrive at the linear problem $A\{\alpha\} = R$.
{The kernels $\kerne_{n\ell}$ vary as $\ell^{3/2}$ (see Eq.~[\ref{asymp}]), and because we choose a large range in $\ell$ ($\ell = 30 - 249$ for MDI for instance), the rows of the matrix $A$ defined in equation~(\ref{matA}) will display large contrasts in magnitude. To mitigate attendant numerical issues, we precondition the equation $A\{\alpha\} = R$ by the diagonal matrix $P = \rm{diag} \{\ell^{-3/2}\}$, transforming the problem to $(PAP)(P^{-1}\{\alpha\}) = PR$. We obtain the vector of coefficients $\{\alpha\}$ through a regularized linear least-squares solution}
\begin{equation}
\{\alpha\} = P[(PAP)^T(PAP) + \mu \,{\bf I}]^{-1}(PAP)^T (PR), \label{alphanl}
\end{equation}
{where ${\bf I}$ is the identity matrix and $\mu$ is an additional regularization term. The preconditioning significantly helps to reduce the condition number of the problem, thereby rendering it better posed.} 

\begin{figure}
\begin{center}
\includegraphics[width=\linewidth,clip=]{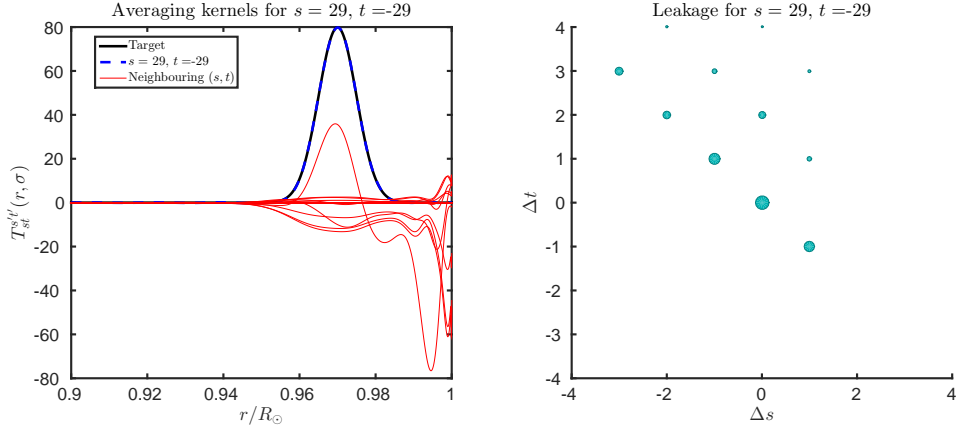}
\end{center}
\caption{
On the left plot, we show an attempt at summing up measurements (and therefore the kernels) to image convective velocities at spatial wavenumber $(s,t) = (29,-29)$ at a depth of $r/R_\odot = 0.97$. The black (thick solid) line is the (desired) target Gaussian. Limited viewing of the Sun results in the inability to isolate individual spherical harmonic oscillation modes, which in turn translates into blurring of the spatial $(s,t)$ domain of perturbations. In this case, we set $\lambda = 0 = \nu$ in equation~(\ref{penalties}), i.e. no leakage penalties. As a consequence, adjacent $(s,t)$ wavenumbers (thin red lines) contribute substantially to the desired inference at $(s,t) = (29,-29)$ (thick blue dashed line; indistinguishable from the black solid line). If the Sun were to be fully visible, the thin red lines would vanish and we would only see the the dashed blue and target Gaussian lines. On the right plot, the leakage is shown pictorially; the big blue dot at $(\Delta s, \Delta t) = (0,0)$ corresponds to the energy associated with the desired (29,-29) mode whereas all other dots denote leakage from adjacent wavenumbers. Indeed, full-Sun coverage would result in an empty plot with a single dot in the centre. Note that the problem of leakage appears to be less severe in this case than in Figure~\ref{leak15noreg}, since the constraint $|t+\Delta t|\le s + \Delta s$ limits the number of modes into which leakage can occur.
}
\label{leak29noreg}
\end{figure}

\begin{figure}
\begin{center}
\includegraphics[width=\linewidth,clip=]{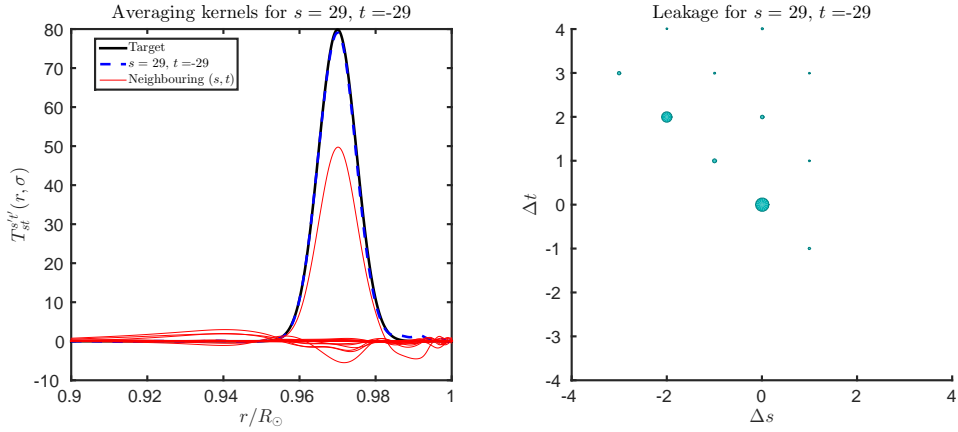}
\end{center}
\caption{
Summing measurements to image convective velocities at spatial wavenumber $(s,t) = (29, -29)$ at a depth of $r/R_\odot = 0.97$ (left). The black solid line is the (desired) target Gaussian. 
For the leakage penalties, we choose $\lambda =0, \nu = 0.5$ in equation~(\ref{penalties}). Therefore the contribution from neighbouring $(s,t)$ wavenumbers (thin red lines) to the desired inference at $(s,t) = (29,-29)$ (the blue dashed line) is much weaker than in Figure~\ref{leak29noreg}. 
Note that since $|t+\Delta t|\le s + \Delta s$, the dots do not cover the entire map.
}
\label{leak29}
\end{figure}

\begin{figure}
\begin{center}
\includegraphics[width=\linewidth,clip=]{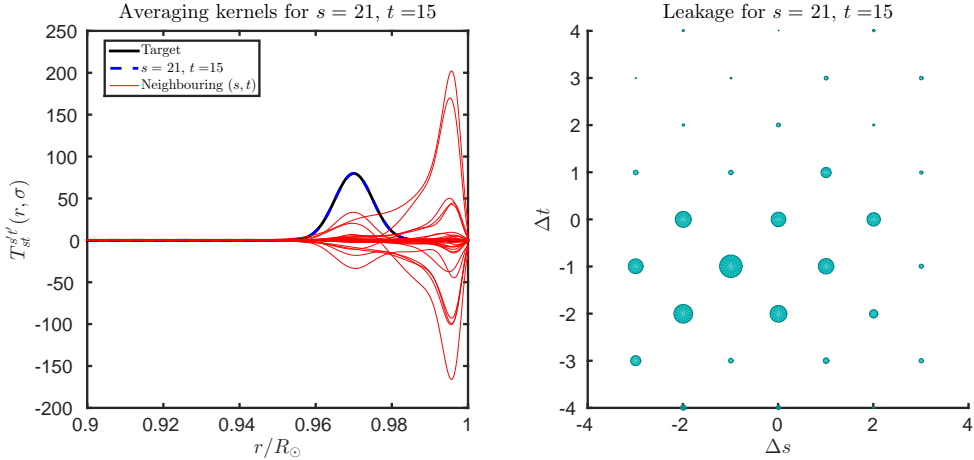}
\end{center}
\caption{
We sum up measurements to image convective velocities at spatial wavenumber $(s,t) = (21,15)$ at a depth of $r/R_\odot = 0.97$ (left panel). The black (thick solid) line is the (desired) target Gaussian. 
As in Figure~\ref{leak29noreg}, we set $\lambda = 0 = \nu$ in equation~(\ref{penalties}), i.e. no leakage penalties. Contributions from neighbouring $(s,t)$ wavenumbers (thin red lines) are seen to overwhelm the desired inference at $(s,t) = (21,15)$ (thick blue dashed line; indistinguishable from the black solid line). 
With penalties described in equation~(\ref{penalties}), the problem of leakage can be greatly mitigated and the inference can be focused on the desired wavenumber and radius, apparent in Figure~\ref{leak15}.
}
\label{leak15noreg}
\end{figure}

\begin{figure}
\begin{center}
\includegraphics[width=\linewidth,clip=]{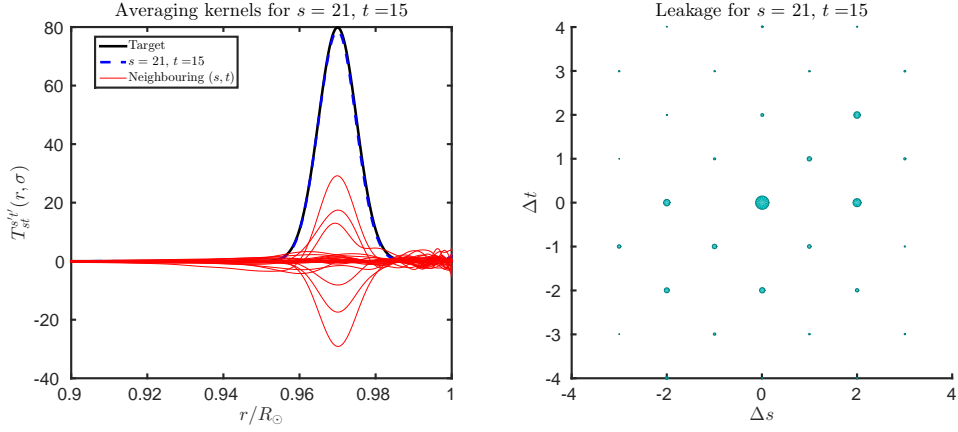}
\end{center}
\caption{
Summing measurements to image convective velocities at spatial wavenumber $(s,t) = (21,15)$ at a depth of $r/R_\odot = 0.97$ (left). 
In contrast to Figure~\ref{leak15}, we apply leakage penalties by setting $\lambda = 0, \nu=0.5$ in equation~(\ref{penalties}). While a variety of neighbouring $(s,t)$ wavenumbers (thin red lines) contribute to the desired inference at $(s,t) = (21,15)$ (thick blue dashed line), they all show power in the desired depth range and there is reduced surface sensitivity, as was the case in Figure~\ref{leak29}. 
}
\label{leak15}
\end{figure}

The primary effect of not penalizing leakage in the $(s,t)$ space as described in equation~(\ref{penalties})
is shown in Figures~\ref{leak29noreg} and~\ref{leak15noreg}. A naive sum over $B$ coefficients shows dominant surface sensitivity to neighbouring $(s,t)$ wavenumbers and weak sensitivity to the wavenumber of interest. In particular, the (undesired) neighbouring-wavenumber sensitivity overwhelms the desired $(s,t) = (21,15)$ sensitivity, thereby rendering interpretation infeasible. This is not altogether surprising since kernels for all helioseismic measurements have strong surface sensitivity \citep[the `shower-glass effect', as described by][]{schunker05} and it typically requires careful analysis to remove the surface contribution. The penalties in equation~(\ref{penalties}) accomplish the tasks of limiting the sensitivity to neighbouring wavenumbers ($\lambda$ term) and the overall surface sensitivity ($\nu$ term). We find empirically that $\lambda=0$ with finite $\nu$ works well for imaging relatively deep layers (i.e. $r/R_\odot < 0.98$ for instance) whereas when imaging the immediate sub-surface, i.e. $r/R_\odot = 0.99$, we set $\nu = 0$ and apply a finite $\lambda$. We show cases for imaging at the depth $r/R_\odot = 0.97$ for two different wavenumbers in Figures~\ref{leak29} and~\ref{leak15}.

Because power from neighbouring wavenumbers leak into the channel we seek to infer, we normalize the $\alpha^{n\ell}$ coefficients thus
\begin{equation}
\alpha^{n\ell} \rightarrow \frac{\alpha^{n\ell}}{\sqrt{\sum_{s't'}\left(\int_\odot dr\,T^{s't'}_{st}\right)^2}}.
\end{equation}

Finally, the average flow component $w_{st}(r_0, \sigma)$ is given by
\begin{equation}
{{{w}}}_{st}(r_0,\sigma) \approx \int_\odot dr\,T^{st}_{st}(r,\sigma)\,w_{st}(r,\sigma) =  {\sum_{n\ell} \alpha^{nl} B_{st}(n,\ell,\sigma)}.
\end{equation}

\section{Noise Model}\label{noise.model}
Correlated realization noise forms a significant fraction of measurement. Accurate inference of the underlying convective flows requires that we model the contribution from correlated noise to $B$ coefficients. 
The analysis may be extended to correlations across different harmonic degrees as well. Following the notation of \citet{woodard16}, 
\begin{equation}
B^\sigma_{st}(\ell) = \frac{\sum_{m,\omega} G^{st*}_{\ell m\omega}\,\phi^{\omega*}_{\ell m}\,\phi^{\omega+\sigma}_{\ell m+t}}{\sum_{m,\omega} |G^{st}_{\ell m\omega}|^2},\label{woodef}
\end{equation}
where $G$ is defined thus
\begin{equation}
G^{st}_{\ell m\omega} = \leak^{\ell m}_{\ell m}\, \leak^{\ell m+t}_{\ell m+t}\,H^\sigma_{\ell\ell mt}\,\gamma^{\ell s\ell}_{tm}.
\end{equation}
Note that this is the same as in equation~(\ref{Bcoef}), except that only diagonal leakage terms are now included.
We model $\phi^\omega_{\ell m}$ as a multi-variate Gaussian random process with the following property \citep[e.g.][]{woodard,gizon_04},
\begin{equation}
\langle\phi^{\omega^*}_{\ell m}\,\phi^{\omega'}_{\ell m+t}\rangle = \delta_{\omega,\omega'} \langle\phi^{\omega^*}_{\ell m}\,\phi^{\omega}_{\ell m+t}\rangle.\label{modphiphi} 
\end{equation}

Let us first consider the contribution of pure noise to the $B$ coefficient by studying the expectation value of equation~(\ref{woodef}),
\begin{equation}
\langle B^\sigma_{st} \rangle = \frac{\sum_{m,\omega} G^{st*}_{\ell m\omega}\,\langle\phi^{\omega*}_{\ell m}\,\phi^{\omega+\sigma}_{\ell m+t}\rangle}{\sum_{m,\omega} |G^{st}_{\ell m\omega}|^2} = \frac{\sum_{m,\omega} G^{st*}_{\ell m\omega}\, \,\langle\phi^{\omega^*}_{\ell m}\,\phi^{\omega}_{\ell m+t}\rangle\delta_{\sigma0}}{\sum_{m,\omega} |G^{st}_{\ell m\omega}|^2},\label{bexp}
\end{equation}
which is therefore trivial except at $\sigma = 0$. Now let us study the second moment (i.e. the variance) of $B$
\begin{equation}
\varepsilon^{st\sigma}_{n\ell}=\langle | B^\sigma_{st}|^2 \rangle = \frac{\sum_{m,m',\omega,\omega'} G^{st*}_{\ell m\omega}\,G^{st}_{\ell m'\omega'}\,\langle\phi^{\omega*}_{\ell m}\,\phi^{\omega+\sigma}_{\ell m+t}\,\phi^{\omega'}_{\ell m'}\,\phi^{\omega'+\sigma*}_{\ell m'+t}\rangle}{[\sum_{m,\omega} |G^{st}_{\ell m\omega}|^2]^2},\label{quart}
\end{equation}
{where we emphasize that $\varepsilon^{st\sigma}_{n\ell}=\langle | B^\sigma_{st}|^2 \rangle$ refers to $B$ coefficients obtained from pure noise, i.e. there being an absence  correlation due to flows or perturbations.}
Isserlis' theorem \citep{isserlis} states that the following relationship governs fourth moments of multivariate zero-mean Gaussian processes $X_i$:
\begin{equation}
\langle X_1\,X_2\,X_3\,X_4\rangle = \langle X_1\,X_2\rangle\langle X_3\,X_4\rangle + \langle X_1\,X_3\rangle\langle X_2\,X_4\rangle + \langle X_1\,X_4\rangle\langle X_2\,X_3\rangle.\label{isserlis}
\end{equation}
Applying equation~(\ref{isserlis}) to~(\ref{quart}), we arrive at
\begin{eqnarray}
\varepsilon^{st\sigma}_{n\ell}={\left[\sum_{m,\omega} |G^{st}_{\ell m \omega}|^2\right]^{-2}}
\sum_{m,m',\omega,\omega'} G^{st*}_{\ell m\omega}\,G^{st}_{\ell m'\omega'}\,(\langle\phi^{\omega*}_{\ell m}\,\phi^{\omega+\sigma}_{\ell m+t}\rangle\langle\phi^{\omega'}_{\ell m'}\,\phi^{\omega'+\sigma*}_{\ell m'+t}\rangle +  \nonumber\\
\langle\phi^{\omega*}_{\ell m}\,\phi^{\omega'}_{\ell m'}\rangle\langle \phi^{\omega+\sigma}_{\ell m+t}\,\phi^{\omega'+\sigma*}_{\ell m'+t}\rangle + 
\langle\phi^{\omega*}_{\ell m}\,\phi^{\omega'+\sigma*}_{\ell m'+t}\rangle\langle\phi^{\omega'}_{\ell m'}\,\phi^{\omega+\sigma}_{\ell m'+t}\rangle ).\nonumber
\end{eqnarray}
The first term resembles equation~(\ref{bexp}) and therefore only contributes at $\sigma=0$, which is of no interest. The second term is
\begin{equation}
\langle\phi^{\omega*}_{\ell m}\,\phi^{\omega'}_{\ell m'}\rangle\langle \phi^{\omega+\sigma}_{\ell m+t}\,\phi^{\omega'+\sigma*}_{\ell m'+t}\rangle = \delta_{\omega,\omega'} \langle\phi^{\omega*}_{\ell m}\,\phi^{\omega}_{\ell m'}\rangle\langle \phi^{\omega+\sigma}_{\ell m+t}\,\phi^{\omega+\sigma*}_{\ell m'+t}\rangle.
\end{equation}
The third term, like the first, is also trivial since over the summation interval $\omega, \omega' > 0$, $\langle\phi^{\omega'}_{\ell m'}\,\phi^{\omega+\sigma}_{\ell m+t}\rangle = 0$.

The noise model for the variance of $B$ coefficients is therefore
\begin{eqnarray}
\varepsilon^{st\sigma}_{n\ell}= \frac{\sum_{m,m',\omega} G^{st*}_{\ell m\omega}\,G^{st}_{\ell m'\omega}\,\langle\phi^{\omega*}_{\ell m}\,\phi^{\omega}_{\ell m'}\rangle\langle \phi^{\omega+\sigma}_{\ell m+t}\,\phi^{\omega+\sigma*}_{\ell m'+t}\rangle 
}{[\sum_{m,\omega} |G^{st}_{\ell m\omega}|^2]^2}.
\end{eqnarray}
Considering the impact of leakage, we have
\begin{eqnarray}
\langle\phi^{\omega*}_{\ell m}\,\phi^{\omega}_{\ell m'}\rangle = \sum_{\elp p\,\eldp p'} \leak^{\elp p}_{\ell m}\,\leak^{\eldp p'}_{\ell m'}\, \langle a^{\omega*}_{\ell' p}\, a^{\omega*}_{\eldp p'}\rangle  = \sum_{\elp p\, \eldp p'} \leak^{\elp p}_{\ell m}\,\leak^{\eldp p'}_{\ell m'}\, N_p\, \delta_{\ell' \eldp}\, \delta_{p' p}\, |R^\omega_{\ell' p}|^2  \nonumber\\
= \sum_{\elp p} \leak^{\elp p}_{\ell m}\,\leak^{\elp p}_{\ell m'}\, N_{\elp}\, |R^\omega_{\ell' p}|^2.
\label{crossphi}
\end{eqnarray}

\subsection{Incomplete model of noise}\label{leaksec}
A fundamental problem with the noise model becomes apparent upon further analysis of equation~(\ref{crossphi}). The observed wavefield $\phi^\omega_{\ell m}$ is complex and so is its correlation. The imaginary and real parts of $\phi^{\omega*}_{\ell m}\,\phi^{\omega}_{\ell m'}$ directly computed from data are comparable in magnitude. However the model on the right-hand side of equation~(\ref{crossphi}) is real since the leakage matrix can be chosen to be real with no loss of generality \citep[e.g.][]{schou94}. The imaginary component of the wavefield correlation would therefore be interpreted as comprising entirely of signal, which is likely not the case.
Thus there are missing ingredients in the model, or perhaps the leakage matrix, and determining what these are exactly is a focus of future work. For the purposes of the present analysis, we make the following assumption,
\begin{equation}
\langle\phi^{\omega*}_{\ell m}\,\phi^{\omega}_{\ell m'}\rangle = \langle|\phi^{\omega}_{\ell m}|^2\rangle\,\delta_{m m'} = \sum_{\elp p} (\leak^{\elp p}_{\ell m})^2 N_{\elp}\, |R^\omega_{\ell' p}|^2,
\end{equation}
which then ensures that two sides of the equation are consistent, giving us the following noise model,
\begin{eqnarray}
\varepsilon^{st\sigma}_{n\ell}= \frac{\sum_{m,\omega} |G^{st}_{\ell m\omega}|^2\langle|\phi^{\omega}_{\ell m}|^2\rangle\,\langle|\phi^{\omega+\sigma}_{\ell m+t}|^2\rangle 
}{[\sum_{m,\omega} |G^{st}_{\ell m\omega}|^2]^2}.\label{final.noise}
\end{eqnarray}

\subsection{Choice of regularization parameter and width of target Gaussian}
\begin{figure}
\begin{center}
\includegraphics[width=\linewidth,clip=]{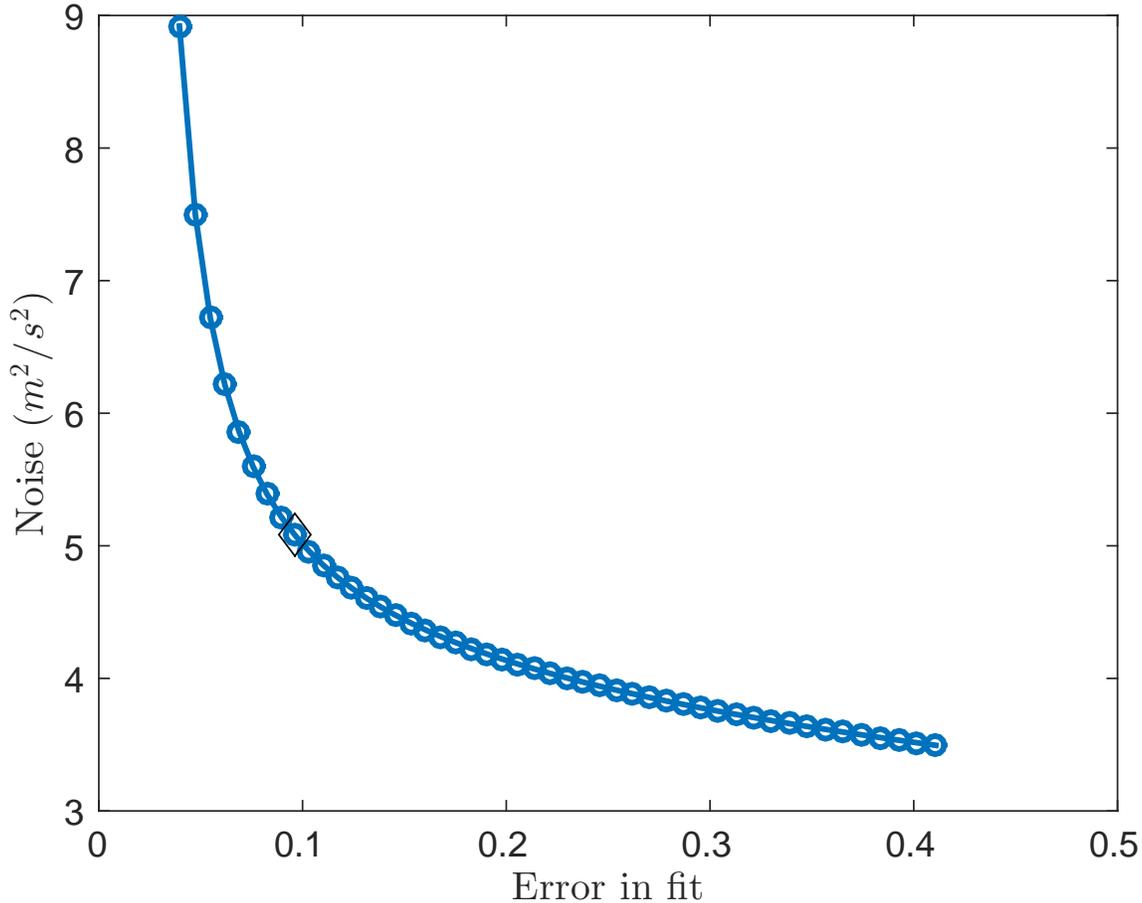}
\end{center}
\caption{{Classical $L$ curve, displaying the tradeoff between goodness of localization in radius (i.e. error in fit) and the amplitude of noise over a range in the regularization parameter $\mu$ (see Eq.~[\ref{defs}]). As the problem is more strongly regularized, the error in obtaining a fit to the desired target function increases whereas at very low regularization, the noise level increases significantly. The knee of the curve, represented by the diamond, corresponds to an optimal choice of $\mu$.}
}
\label{lcurve}
\end{figure}

\begin{figure}
\begin{center}
\includegraphics[width=\linewidth,clip=]{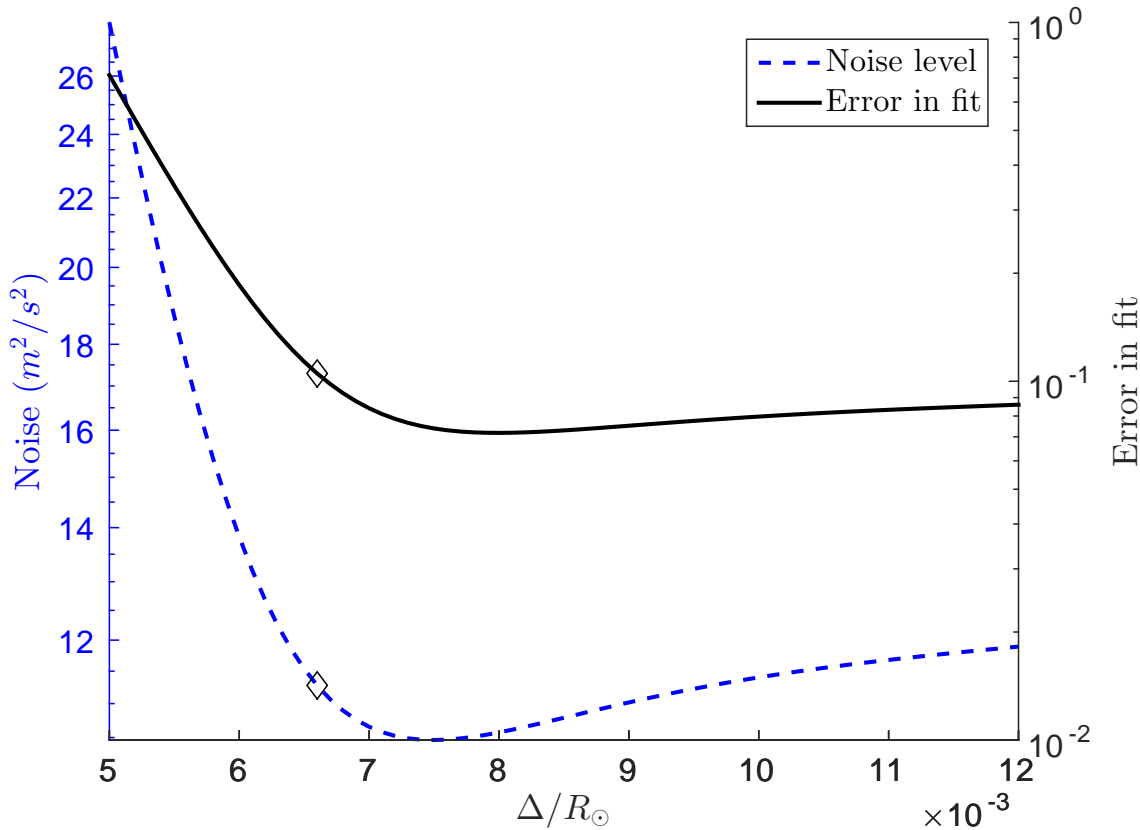}
\end{center}
\caption{{Determining the optimal width $\Delta$ of a target Gaussian function ${\mathcal T}(r) = ({2\pi \Delta^2})^{-1/2}\,\,\exp[-(r-r_0)^2/(2\Delta^2)]$, for the mode $s=29, t = -25$ and centre radius $r_0 = 0.95 R_\odot$. As $\Delta$ becomes smaller, both the error in fit and measurement noise rise (given by Eq.[\ref{defs}]). We choose the optimal point as corresponding to the knee of the noise curve, marked by the diamond, which is $\Delta/R_\odot \approx 0.007$.}
}
\label{tradewidth}
\end{figure}

{In order to set the regularization parameter $\mu$ in equation~(\ref{alphanl}), which controls the tradeoff between the level of noise in the chosen combination of measurements and ability to fit the target function, we plot the error in the fit against the noise for a range of values of $\mu$. The goodness of fit and noise level are described as}
\begin{equation}
{\rm Error\, in\, fit} = \int_\odot dr\, (T^{st}_{st} - \target)^2,\,\,\,\,\,\, {\rm Noise\, level} = \sum_{n\ell}\alpha^2_{n\ell} \varepsilon^{st\sigma}_{n\ell}.\label{defs}
\end{equation}
{The ideal degree of regularization corresponds to the knee of the classical $L$ curve, as shown in Figure~\ref{lcurve}. The width of the target Gaussian function is another parameter that needs to be determined. Although this is a two-parameter problem (i.e. $\Delta$ and $\mu$), we fix the value of $\mu$, finding it to be a robust quantity, and vary the value of $\Delta$, the width of the target Gaussian at a desired depth. Both error in fit and noise fall as $\Delta$ increases, and the smallest value of $\Delta$ that accommodates an acceptable noise level corresponds to the knee of the curve, as shown in Figure~\ref{tradewidth}.}

\subsection{Estimating signal (flow) power}
\begin{figure}
\begin{center}
\includegraphics[width=\linewidth,clip=]{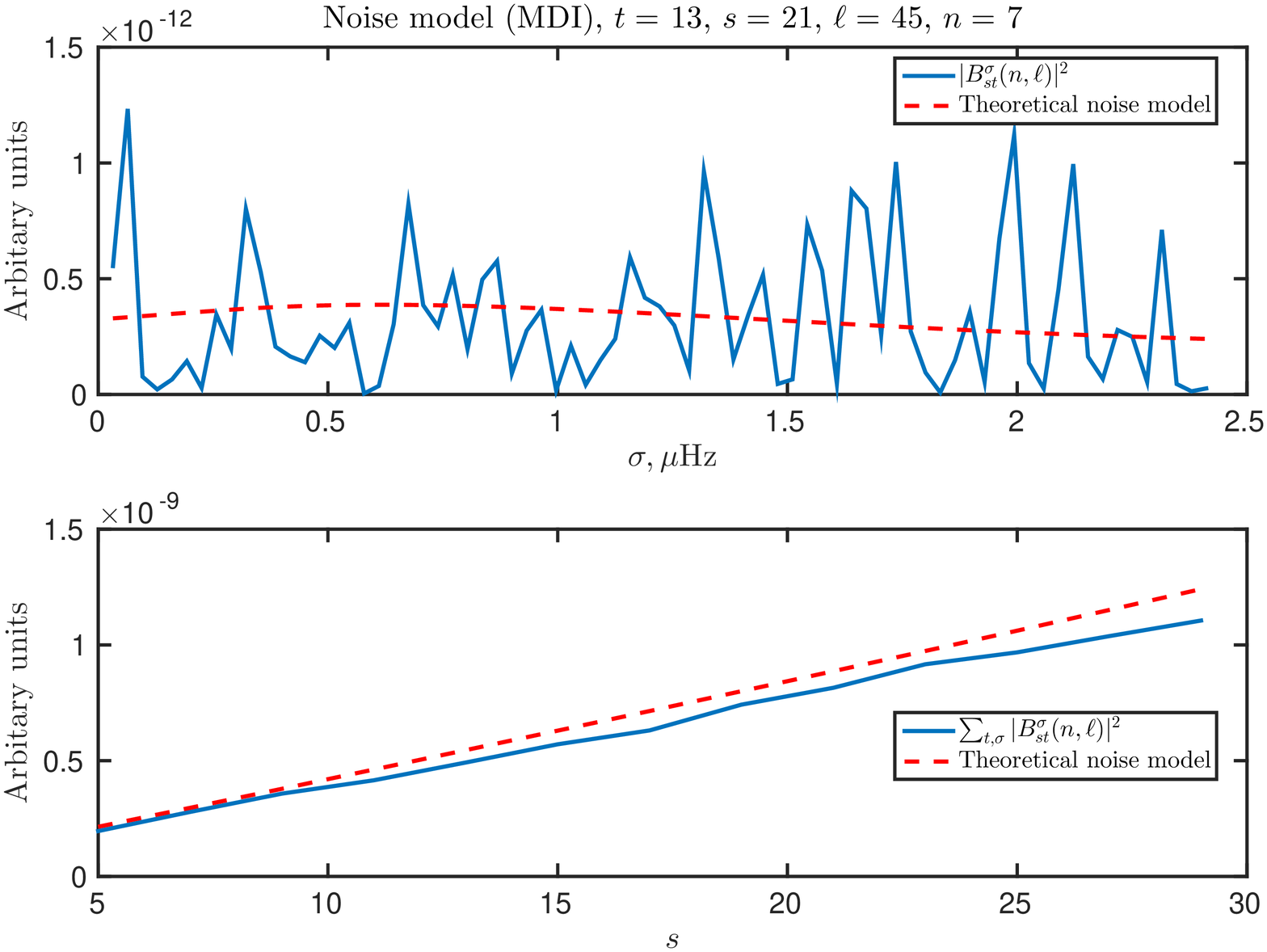}
\end{center}
\caption{
Noise model for Michelson-Doppler-Imager (MDI) measurements of normal-mode coupling $B^\sigma_{st}$ from one year's data, obtained from the $\ell =45, n = 7$ mode. Upper panel: the solid blue line shows $|B^\sigma_{st}|^2$ as a function of frequency $\sigma$ at a specific $(s,t) = (21,13)$ and the dashed red line is the theoretical noise model in equation~(\ref{final.noise}). It is seen that the theoretical noise model captures aspects of the variance of the $|B^\sigma_{st}|^2$ curve although a more thorough study is required to establish this - specifically at high frequencies. Lower panel: Both curves summed over $t$ and $\sigma$ and plotted as a function of harmonic degree $s$. This curve suggests that the power in $|B^\sigma_{st}|^2$ is essentially consistent with realization noise.
}
\label{noise45}
\end{figure}

\begin{figure}
\begin{center}
\includegraphics[width=\linewidth,clip=]{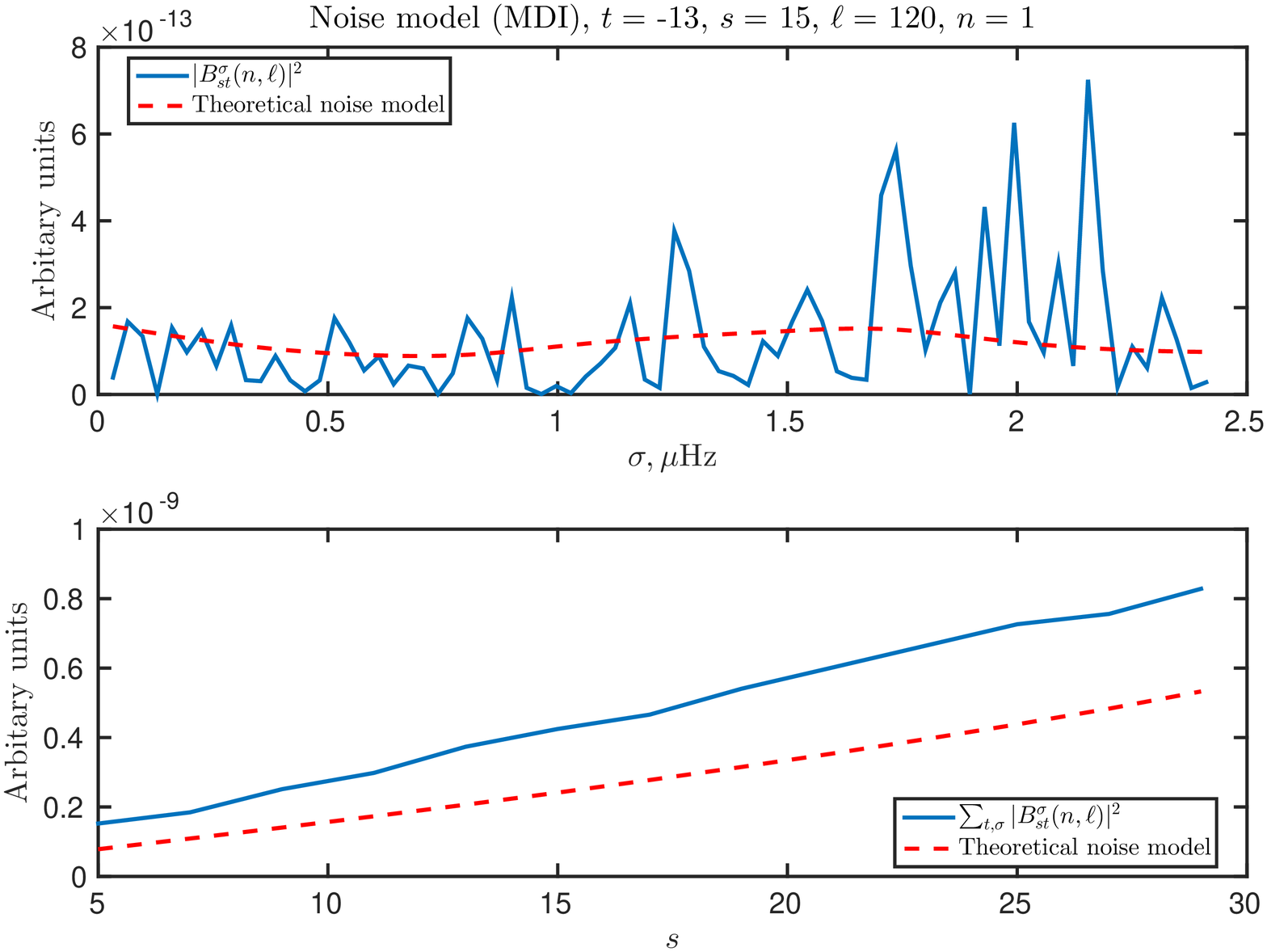}
\end{center}
\caption{
Noise model for MDI measurements of normal-mode coupling $B^\sigma_{st}$ from one year's data, obtained from the $\ell =120, n = 1$ mode. Upper panel: the solid blue line shows $|B^\sigma_{st}|^2$ as a function of frequency $\sigma$ at a specific $(s,t) = (15,-13)$, the dashed red line is the theoretical noise model in equation~(\ref{final.noise}).  Lower panel: The noise and signal summed over $t$ and $\sigma$ and plotted as a function of harmonic degree $s$. In contrast to Figure~\ref{noise45}, it is seen that the theoretical model predicts a much lower noise level than the $|B^\sigma_{st}|^2$ curve, suggesting the existence of signal in this $(s,t)$ channel as sensed by $\ell=120, n= 1$. 
}
\label{noise120}
\end{figure}

\begin{figure}
\begin{center}
\includegraphics[width=\linewidth,clip=]{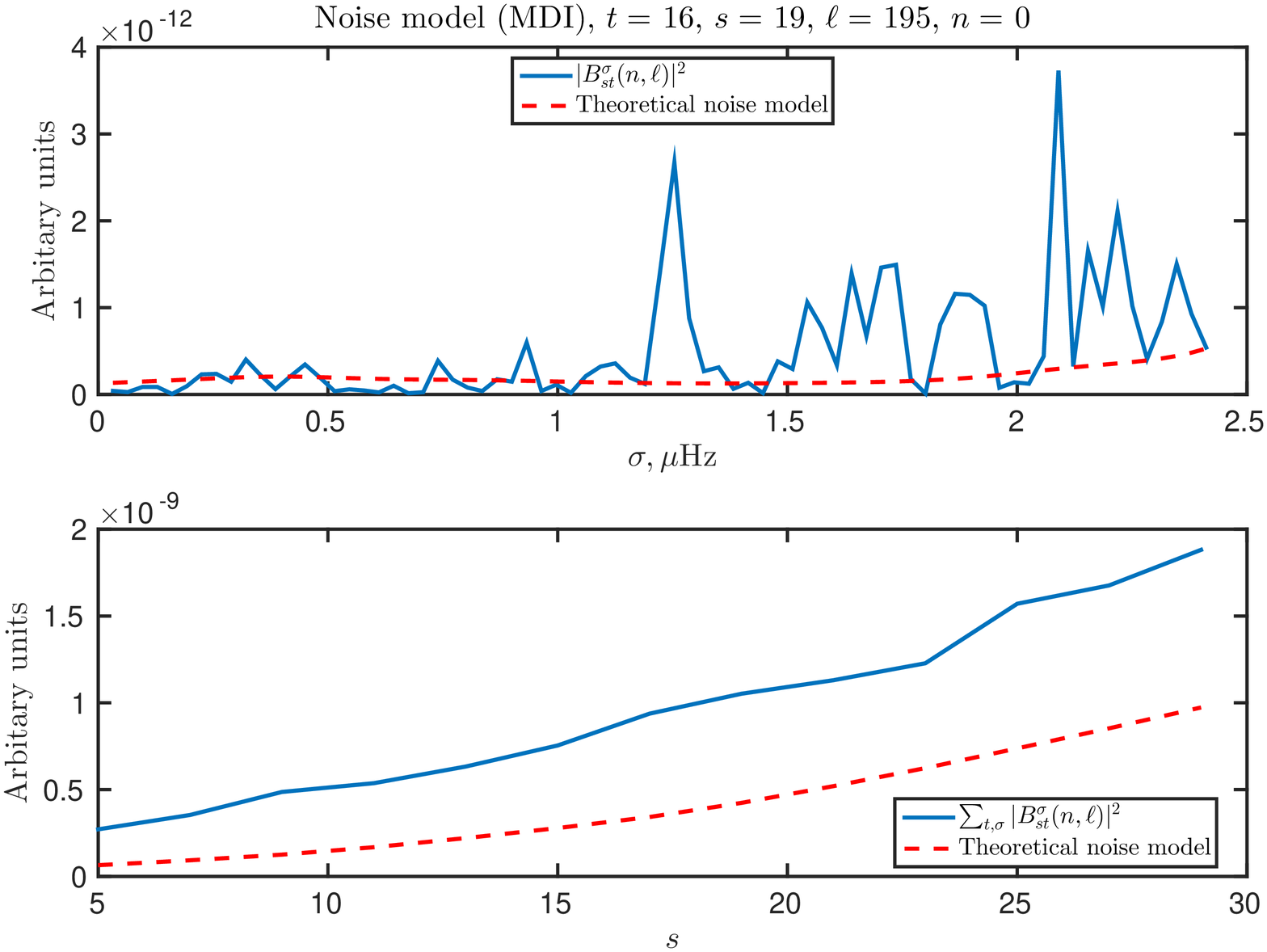}
\end{center}
\caption{
Noise model for MDI measurements of normal-mode coupling $B^\sigma_{st}$ from one year's data, obtained from the $\ell =195, n = 0$ mode. Upper panel: the solid blue line shows $|B^\sigma_{st}|^2$ as a function of frequency $\sigma$ at a specific $(s,t) = (15,-13)$, the dashed red line is the theoretical noise model in equation~(\ref{final.noise}). Lower panel: The noise and signal summed over $t$ and $\sigma$ and plotted as a function of harmonic degree $s$. In contrast to Figure~\ref{noise45} and similar to Figure~\ref{noise120}, it is seen that the theoretical model predicts a much lower noise level than the $|B^\sigma_{st}|^2$ curve, suggesting the existence of signal in this $(s,t)$ channel as sensed by $\ell = 195, n=0$. 
}
\label{noise195}
\end{figure}

The analysis set out in Section~\ref{SOLA} along with this noise model describes how to subtract noise from B-coefficients, i.e.
\begin{equation}
{\rm signal} =  \left| \sum_{n,\ell} \alpha^{n\ell}\, B^\sigma_{st}(n,\ell) \right|^2  - \sum_{n,\ell}\alpha^2_{n\ell} \varepsilon^{st\sigma}_{n\ell},\label{signoise}
\end{equation}
where the left-hand side is the signal, the first term on the right-hand side represents the squared absolute value of the weighted sum of the observed $B$ coefficients and the second term is the weighted sum of the expected noise for each coefficient. Note that the variance in the power (`noise of the noise') quantifies the expected error bars on the inferences of the signal. This requires evaluating an eighth moment of the wavefield, and the calculation is outlined in appendix~\ref{varnoise}. At present, we have not evaluated these terms. In Figures~\ref{noise45},~\ref{noise120} and~\ref{noise195}, we plot the $B$-coefficient and noise power as functions of temporal frequency $\sigma$ and harmonic degree $s$. The observed $B$-coefficients were computed for specific oscillation modes (described in the attendant captions) from 1 year of Michelson Doppler Imager observations \citep[1996 -1997;][]{scherrer95}. In Figure~\ref{noise45}, the noise model and $B$-coefficient power are essentially consistent with each other whereas in Figures~\ref{noise120} and~\ref{noise195}, $B$-coefficient power is notably larger than the theoretically predicted noise level (Eq.~[\ref{final.noise}]).

\section{Conclusions}
We have studied the problem of how to best combine wavefield-correlation measurements in the Sun in order to obtain optimal inferences of desired properties in the solar interior. These correlations are measured from observations using a linear-least-squares methodology described in e.g. \citet{woodard16}. A great benefit of using this technique is that it is a linear functional of the wavefield, unlike the non-linear least squares approach adopted by e.g. \citet{schou92} to obtain global-mode parameters. There is a downside to adopting a straightforward method, namely that spatial leakage between oscillation modes is transmitted directly to a leakage between wavenumbers of the perturbation. Thus it might ostensibly be difficult to isolate individual spherical-harmonic wavenumbers of the perturbation. As a first step, we derive the sensitivities of the normal-mode-coupling measurement (specifically the so-called $B$ coefficient) to perturbations, finding that, in addition to being dominantly sensitive to $(s,t)$ it is also sensitive to adjacent wavenumbers, i.e. within the range $(s\pm2,t\pm2)$. The question of how to optimally combine the measured $B$ coefficients such that it is overwhelmingly sensitive only to the $(s,t)$ wavenumber while being sharply localized at a desired depth (and no surface tail) and also ensuring that noise level of the inferred flow velocity is reasonable. To that end, we build a noise model for $B$ coefficients, which can then be propagated to the inferred flow velocities. We find (to our pleasant surprise) that when combining a sufficient number of modes, all these constraints are well met. The issue of leakage suggests that the convective-velocity-amplitude estimates of \citet{woodard16}, which were obtained by direct sums over harmonic degree, are associated with surface layers (as opposed to the interior) and may therefore explain why they are relatively large. 
Overall, the present results strongly encourage a more complete exploration of the possibilities of using normal-mode coupling to infer the internal properties of the Sun.


\begin{acknowledgments}
SMH acknowledges support from Ramanujan Fellowship SB/S2/RJN-73 and the Max-Planck Partner group program. SMH is grateful to Martin Woodard, with whom many useful conversations were had, for helping resolve several technical issues. This research was supported in part by the International Centre for Theoretical Sciences (ICTS) during a visit for participating in the program - Turbulence from Angstroms to Light Years (ICTS/Prog-taly2018/01).
\end{acknowledgments}

\bibliographystyle{apj}
\bibliography{ms.bbl}

\appendix

\section{Derivation of asymptotic kernels}\label{asympsec}
\begin{figure}
\begin{center}
\includegraphics[width=\linewidth,clip=]{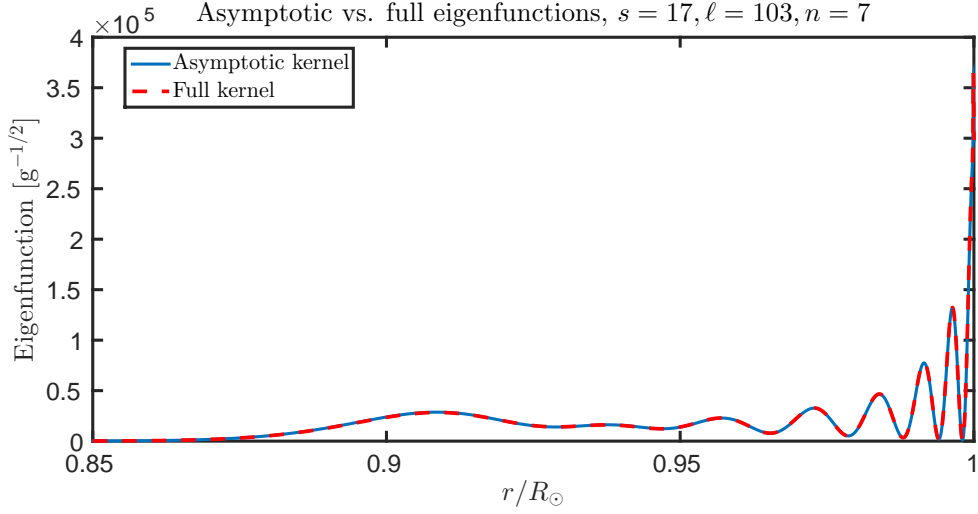}
\end{center}
\caption{
Comparison between exact (Eq.~[\ref{truekern}]) and asymptotic (Eq.~[\ref{asymp}]) kernels for $\ell = 103, n = 7, s = 17$ plotted as a function of radius. It is seen that these are essentially indistinguishable. Although not shown here, we find that the asymptotic and full expressions for these kernels are remarkably well matched over broad ranges of $n, \ell, s$. Note that the comparison gets slightly worse when $s \approx \ell$ but the match even in the worst cases was very good (not shown here). 
}
\label{kernelfig}
\end{figure}

The kernel sensitive to toroidal flows is given by \citep[Eq.~60 of][]{hanasoge17_etal}
\begin{eqnarray}
K^{st}_w = \frac{\rho r\omega_{n\ell}}{2\sqrt{\pi}}\ell'(\ell'+1)\sqrt{(2\ell'+1)(2\ell+1)}\gamma^{\ell's\ell}_{tm}(1 - (-1)^{\ell'+\ell+s}) \begin{pmatrix}\ell' & s & \ell\\ -1 & 0 & 1\end{pmatrix}\times\nonumber\\
(-UU' + VU' + UV' -\frac{VV'}{2} [\ell'(\ell'+1) + \ell(\ell+1) - s(s+1)] ),\label{truekern}
\end{eqnarray}
where $U,V$ are eigenfunctions corresponding to spherical-harmonic degree $\ell$ and similarly for the primed quantities. The phase factor $1-(-1)^{\ell'+\ell+s}$ constrains the kernel to be non-zero only when $\ell' + s + \ell$ is odd. 
We use the following asymptotic relation for the Wigner-3j symbol \citep[appendix A, Eq.~A3 of][]{vorontsov11}
\begin{equation}
\begin{pmatrix}\ell' & s & \ell\\ -m & 0 & m\end{pmatrix} \approx \frac{(-1)^{\ell'+m}}{\sqrt{2\ell}}\left[\frac{(s-\ell'+\ell)!}{(s+\ell'-\ell)!}\right]^{\frac{1}{2}} P^{\ell'-\ell}_{s}\left(\frac{m}{\ell}\right),
\label{asymp1}
\end{equation}
where $P^\ell_{s}$ is the associated Legendre polynomial and the approximation is valid in the limit $s\ll \ell, \ell'$. Further assuming that $\ell\approx\ell' \gg 1$ and considering the specific case of $t=0$ and $m = 1 = m'$, the following approximation is known to apply \citep[appendix A, Eq.~A4 of][]{vorontsov11} when $s + \ell' + \ell$ is odd
\begin{equation}
\begin{pmatrix}\ell' & s & \ell\\ -m & 0 & m\end{pmatrix}\begin{pmatrix}\ell' & s & \ell\\ -1 & 0 & 1\end{pmatrix} \approx 
(-1)^\frac{s-\ell'+\ell+2m+1}{2}\frac{(s-\ell'+\ell)!! (s+\ell'-\ell)!!}{2\ell^2(s+\ell'-\ell)!} P^{\ell'-\ell}_s\left(\frac{m}{\ell}\right).\label{asymp2}
\end{equation}
Dividing asymptotic relation~(\ref{asymp2}) by~(\ref{asymp1}), we arrive at
\begin{equation}
\begin{pmatrix}\ell' & s & \ell\\ -1 & 0 & 1\end{pmatrix} \approx 
(-1)^{\frac{s-\ell'+\ell+1}{2} - \ell'}\frac{(s-\ell'+\ell)!! (s+\ell'-\ell)!!}{\sqrt{2}\ell^\frac{3}{2}\sqrt{(s+\ell'-\ell)!(s-\ell'+\ell)!}}.\label{asymp.exp}
\end{equation}
Approximating $U'\approx U, V'\approx V, |UV| \ll |U^2|, \ell(\ell+1) |V^2|$, $\ell'(\ell'+1) +\ell(\ell+1) \gg s(s+1)$ and $\ell\approx\ell'$, and leaving the Wigner-3j symbols alone for now, we obtain
\begin{equation}
K^{st}_w(r) = -\frac{2}{\sqrt{\pi}} \ell^3\,\gamma^{\ell's\ell}_{tm}\rho r\omega_{n\ell} \begin{pmatrix}\ell' & s & \ell\\ -1 & 0 & 1\end{pmatrix}
\left[U^2 + \ell(\ell+1){V^2} \right].\label{asymp.final}
\end{equation}
Substituting the asymptotic expression~(\ref{asymp.exp}) for the Wigner-3j symbol, multiplying equation~(\ref{asymp.final}) by $(-1)^{\ell' + \ell' - \ell + \ell}$, replacing the negative sign by $(-1)^{-1}$ and rearranging, we arrive at 
\begin{equation}
K^{st}_w(r) \approx (-1)^{\frac{s+\ell'-\ell-1}{2}} (-1)^{\ell}{\ell^\frac{3}{2}}{\sqrt{\frac{2}{\pi}}}\gamma^{\ell's\ell}_{tm}\rho r\omega_{n\ell}
\left[U^2+ V^2 \ell(\ell+1) \right] \frac{(s-\ell'+\ell)!! (s+\ell'-\ell)!!}{\sqrt{(s+\ell'-\ell)!(s-\ell'+\ell)!}}.\label{asymp}
\end{equation}
Note that there is a factor of 2 discrepancy between asymptotic expression~(\ref{asymp}) and equations~(45) and~(46) of \citet{woodard14}. To confirm that the expressions derived here are correct, we plot the exact and asymptotic expressions for $\ell = 103, n = 7, s = 17$ in Figure~\ref{kernelfig}.

\section{Variance of the noise}\label{varnoise}
We compute the variance of the noise in order to place error bars on the degree of confidence in bounding the magnitudes of convective velocities. The variance is given by $\langle | B^\sigma_{st}|^4 \rangle - \langle | B^\sigma_{st}|^2 \rangle^2$ and  we need to compute the first term, which is
\begin{equation}
\langle | B^\sigma_{st}|^4 \rangle = \frac{\sum\,G^{st*}_{\ell m\omega}\,G^{st}_{\ell m'\omega'}\,G^{st*}_{\ell \emdp\omdp}\,G^{st}_{\ell \emtp\omtp}\,\langle\phi^{\omega*}_{\ell m}\,\phi^{\omega'}_{\ell m'}\, \phi^{\omega+\sigma}_{\ell m+t}\,\phi^{\omega'+\sigma*}_{\ell m'+t}\phi^{\omdp*}_{\ell\emdp}\,\phi^{\omtp}_{\ell \emtp}\, \phi^{\omdp+\sigma}_{\ell \emdp+t}\,\phi^{\omtp+\sigma*}_{\ell \emtp+t}\rangle 
}{[\sum_{m,\omega} |G^{st}_{\ell m\omega}|^2]^4},\label{eightorder}
\end{equation}
where the sum is applied over repeated indices.
{Invoking Isserlis' theorem, the expectation of the eighth-order product in equation~(\ref{eightorder}) produces 105 terms of products of pair-wise correlations. Most of these terms disappear when using the correlation model given in equation~(\ref{modphiphi}) and restricting summations only to positive intervals in frequency.}
Upon simplification, we obtain the following twelve terms
\begin{eqnarray}
\langle | B^\sigma_{st}|^4 \rangle = {\left[\sum_{m,\omega} |G^{st}_{\ell m\omega}|^2\right]^{-4}}\left[2\left(\sum_{m,\omega}  |G^{st}_{\ell m\omega}|^2\,P^\omega_{\ell m}\,P^{\omega+\sigma}_{\ell m+t}\right)^2 +  2\sum_{m,\omega}  |G^{st}_{\ell m\omega}|^4\,\left(P^{\omega}_{\ell m}\,P^{\omega+\sigma}_{\ell m+t}\right)^2 + \right. \nonumber\\
\left. \sum_{m,\omega} 2 |G^{st}_{\ell m\omega}|^2\,|G^{st}_{\ell m-t,\omega-\sigma}|^2\,P^{\omega-\sigma}_{\ell m-t} \,\left(P^{\omega}_{\ell m}\right)^2\,P^{\omega+\sigma}_{\ell m+t} +  |G^{st}_{\ell m\omega}|^2\,|G^{st}_{\ell m+t\omega+\sigma}|^2\,\left(P^{\omega}_{\ell m}\right)^2\,P^{\omega+\sigma}_{\ell m+t}\,P^{\omega+2\sigma}_{\ell m+2t} \right.\nonumber\\
\left. + 
|G^{st}_{\ell m\omega}|^2\,|G^{st}_{\ell m+t,\omega+\sigma}\,G^{st}_{\ell m-t,\omega-\sigma}|\,\left(P^{\omega+\sigma}_{\ell m+t}\right)^2\,P^{\omega+2\sigma}_{\ell m+2t}\,P^{\omega}_{\ell m} \right],\nonumber\\
%
\end{eqnarray}
where we define 
$P^\omega_{\ell m} = \langle|\phi^\omega_{\ell m}|^2\rangle$ and $P^{\omega+\sigma}_{\ell m + t} = \langle|\phi^{\omega+\sigma}_{\ell m+t}|^2\rangle$ and so on. We do not compute this term in the present analysis.


\section{Mode normalization constant}
The power spectrum of an oscillation mode is given by a Lorentzian \citep[e.g.][]{schou94},
\begin{equation}
P(\omega) = \frac{P_{\rm tot}{\Gamma_{n\ell}/2}}{ {(\omega - \omega_{n\ell})^2}+{(\Gamma_{n\ell}/2)^2}},\label{eqpow1}
\end{equation}
and defining $P_{\rm tot} = C_{n\ell} A_{n\ell}^2 / 2$, where $A_{n\ell}$ is the measured amplitude of the mode in question and $C_{n\ell}$ is an overall normalization factor described below.
From the definition of the power spectrum,
\begin{equation}
\langle |\phi^\omega_{\ell m}|^2\rangle = N_\ell\,\sum_{\ell' m'} (L^{\ell' m'}_{\ell m})^2 |R^\omega_{\ell' m'}|^2,\label{powspec}
\end{equation}
where $R^\omega_{\ell m}$ stated in equation~(\ref{Rdef}). Thus the power spectrum is given by
\begin{equation}
\langle |\phi^\omega_{\ell m}|^2\rangle \approx \frac{N_{\ell}}{4\omega_{n\ell}^2}\sum_{\ell' m'} \frac{(L^{\ell' m'}_{\ell m})^2}{(\omega_{n\ell' m'} - \omega)^2 + (\Gamma_{n\ell'}/2)^2 }.\label{eqpow2}
\end{equation}

Comparing equations~(\ref{eqpow1}) and~(\ref{eqpow2}), we obtain $N_\ell = C_{n\ell} A^2_{n\ell} \omega^2_{n\ell} \Gamma_{n \ell}$, where $C_{n\ell}$ is defined as
\begin{equation}
C_{n\ell} = \frac{\sum_{m,\omega} |\phi^\omega_{\ell m}|^2}{A^2_{n\ell} \omega^2_{n\ell} \Gamma_{n \ell}\sum_{m,\omega}\sum_{\ell',m'} (L^{\ell' m'}_{\ell m})^2[(\omega_{n\ell' m'} - \omega)^2 + (\Gamma_{n\ell'}/2)^2]^{-1}}.
\end{equation}
The numerator and denominator are summed over all $m$ and $\omega$ in the $m$-dependent interval $\omega \in (\omega_{n\ell m} -2\Gamma_{n\ell},\omega_{n\ell m} -2\Gamma_{n\ell})$. This provides an overall normalization constant to be applied at each radial order and harmonic degree. 
Note that because we only consider coupling at the same radial order, we drop the subscript $n$ on the mode normalization constant $N_\ell$. 



\section{Surface-tail constraint}\label{surfconstraint}
In the expression~(\ref{penalties}), we apply two penalties ($\lambda$ and $\nu$), the former to diminish sensitivity to undesired wavenumbers and the latter to the overall surface tail. Since modes spend a lot of time near the surface layers, kernels typically tend to be dominantly sensitive to surface layers. Only when taking a coherent weighted sum of the kernels can the surface sensitivity be removed. However, in the present case, numerous wavenumbers contribute to the overall sensitivity. Weights designed for a specific $(s,t)$ that result in sensitivity at depth, for instance, do not produce the same depth sensitivity in adjacent $(s',t')$ wavenumbers (as in Figures~\ref{leak29noreg} through~\ref{leak15}). In fact, because the sum is not necessarily coherent in the same manner, the sensitivity of the adjacent wavenumbers is to the surface. We can therefore penalize sensitivity to the surface layers as an indirect means of mitigating leakage in the $(s,t)$ space. Note that this doesn't apply when imaging the immediate sub-surface layers, and so we set $\nu = 0$ in that case, and choose finite $\lambda$ instead in equation~(\ref{penalties}).

\begin{figure}
\begin{center}
\includegraphics[width=0.5\linewidth,clip=]{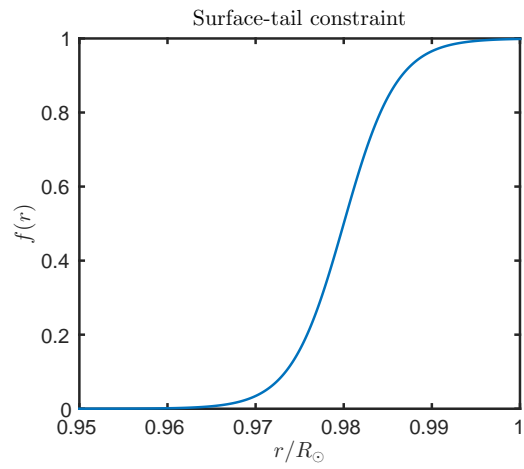}
\end{center}
\caption{
Surface-tail constraint function $f(r)$, as used in equation~(\ref{penalties}) and described in equation~(\ref{surfeq}). The function is zero for most of the interior and rises sharply in the outer 2.5\% of the solar radius or so.
}
\label{surfcontrol}
\end{figure}

For the function $f(r)$, we choose the following
\begin{equation}
f(r) = \frac{1}{1 + \exp\frac{0.98 R_\odot - r}{0.003 R_\odot}},\label{surfeq}
\end{equation}
plotted in Figure~\ref{surfcontrol}.

\end{document}